\pdfoutput=1
\documentclass[aps,prl,twocolumn,showpacs,superscriptaddress,a4paper,
    floatfix]{revtex4}
\usepackage[utf8]{inputenc}
\usepackage{graphicx}
\usepackage{amsmath,amssymb}
\usepackage{hyperref}

\begin{document}

\title{
  Quantitative Determination of Temperature in the Approach to
  Magnetic Order of Ultracold Fermions in an Optical Lattice
}

\author{R. J\"ordens}
\affiliation{Department of Physics, ETH Zurich, 8093 Zurich,
Switzerland}

\author{L. Tarruell}
\affiliation{Department of Physics, ETH Zurich, 8093 Zurich,
Switzerland}

\author{D. Greif}
\affiliation{Department of Physics, ETH Zurich, 8093 Zurich,
Switzerland}

\author{T. Uehlinger}
\affiliation{Department of Physics, ETH Zurich, 8093 Zurich,
Switzerland}

\author{N. Strohmaier}
\affiliation{Department of Physics, ETH Zurich, 8093 Zurich,
Switzerland}

\author{H. Moritz}
\affiliation{Department of Physics, ETH Zurich, 8093 Zurich,
Switzerland}

\author{T. Esslinger}
\email{esslinger@phys.ethz.ch}
\affiliation{Department of Physics,
ETH Zurich, 8093 Zurich, Switzerland}

\author{L. De Leo}
\affiliation{Centre de Physique Th\'eorique, CNRS, \'Ecole Polytechnique,
91128 Palaiseau Cedex, France}

\author{C. Kollath}
\affiliation{Centre de Physique Th\'eorique, CNRS, \'Ecole Polytechnique,
91128 Palaiseau Cedex, France}

\author{A. Georges}
\affiliation{Centre de Physique Th\'eorique, CNRS, \'Ecole Polytechnique,
91128 Palaiseau Cedex, France}
\affiliation{Coll\`ege de France, 11 place Marcelin Berthelot,
75231 Paris Cedex, France}

\author{V. Scarola}
\affiliation{Department of Physics, Virginia Tech, Blacksburg,
Virginia 24061, USA}

\author{L. Pollet}
\affiliation{Department of Physics, Harvard University, Cambridge,
Massachusetts 02138, USA}

\author{E. Burovski}
\affiliation{LPTMS, CNRS and Universit\'e Paris-Sud, UMR8626, B\^at.~100,
91405 Orsay, France}

\author{E. Kozik}
\affiliation{Department of Physics, ETH Zurich, 8093 Zurich,
Switzerland}

\author{M. Troyer}
\affiliation{Department of Physics, ETH Zurich, 8093 Zurich,
Switzerland}

\date{\today}

\begin{abstract}

  We perform a quantitative simulation of the repulsive Fermi-Hubbard model
  using an ultra-cold gas trapped in an optical lattice. The entropy of the
  system is determined by comparing accurate measurements of the
  equilibrium double occupancy with theoretical calculations over a wide
  range of parameters.  We demonstrate the applicability of both
  high-temperature series and dynamical mean-field theory to obtain
  quantitative agreement with the experimental data.  The reliability of
  the entropy determination is confirmed by a comprehensive analysis of all
  systematic errors.  In the center of the Mott insulating cloud we obtain
  an entropy per atom as low as $0.77k_\text{B}$ which is about twice as
  large as the entropy at the N\'eel transition. The corresponding
  temperature depends on the atom number and for small fillings reaches
  values on the order of the tunneling energy.

\end{abstract}


\maketitle

Experimental progress in the field of atomic quantum gases has led to a new
approach to quantum many-body physics. In particular, the combination of
quantum degenerate and strongly interacting Fermi gases \cite{DeMarco1999,
Loftus2002} with optically induced lattice potentials \cite{Greiner2002}
now allows the study of a centerpiece of quantum condensed matter physics,
the Fermi-Hubbard model \cite{Koehl2005}. The high level of control over
the atomic systems has led to the concept of quantum simulation, which for
the case of the Fermi-Hubbard model is expected to provide answers to
intriguing open questions of frustrated magnetism and $d$-wave
superfluidity \cite{Hofstetter2002}. Recent experiments \cite{Jordens2008,
Schneider2008} have indeed demonstrated that the strongly correlated regime
of the repulsive Fermi-Hubbard model is experimentally accessible and the
emergence of a Mott insulating state has been observed. In this Letter, we
succeed in performing a quantitative simulation of the Fermi-Hubbard model
using cold atoms. The level of precision of the experiment enables us to
determine the entropy and the temperature of the system, and thereby to
quantify the approach to the low temperature phases.

The main challenge for the quantum simulation of the Fermi-Hubbard model is
a further reduction in temperature. Here the lack of a quantitative
thermometry method in the lattice is a key obstacle.  For strongly
correlated bosonic systems thermometry has recently been demonstrated by
direct comparison with quantum Monte-Carlo simulations \cite{Trotzky2009}
or by using the boundary of two spin polarized clouds \cite{Weld2009}. In
the fermionic case, previous methods to determine the temperature could be
used only in limiting regimes of the Hubbard model, namely the
noninteracting \cite{Koehl2006, Strohmaier2007} and zero-tunneling
\cite{Jordens2008, Scarola2009} regimes. However, intermediate interactions
are most interesting for quantum simulation of the Fermi-Hubbard model and
no reliable thermometry method has been available up to now.

\begin{figure}[!Htb]
  \includegraphics{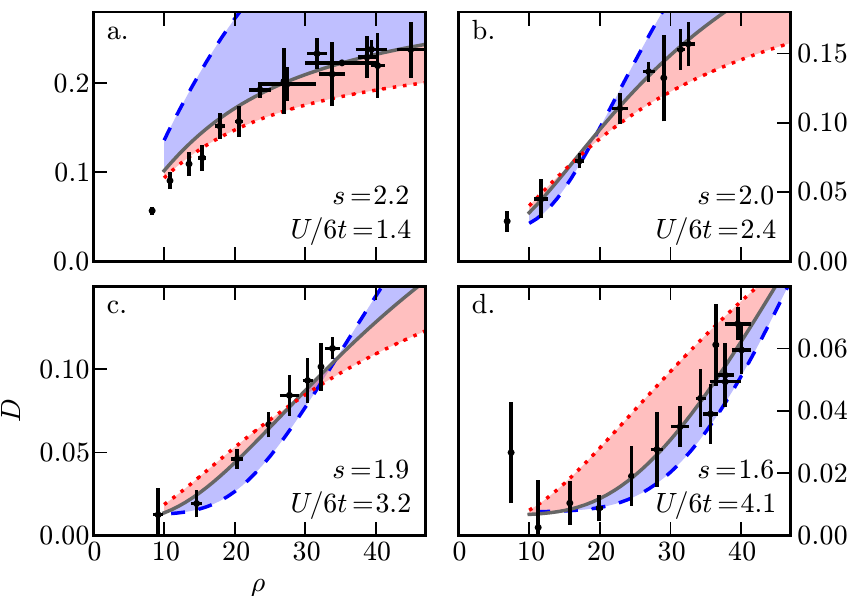}

    \caption{Double occupancy: experiment versus theory. Points and error
    bars are the mean and standard deviation of at least three experimental
    runs. The solid curve in each panel is the best fit of the second order
    high-temperature series to the experimental data and yields specific
    entropies of $s=2.2(2), 2.0(5), 1.9(4), 1.6(4)$ for the different
    interactions strengths of $U/6t=1.4(2), 2.4(4), 3.2(5), 4.1(7)$.
    Curves for $s=1.3$ (dashed curve) and $2.5$ (dotted curve) represent
    the interval of specific entropy measured before and after the ramping
    of the lattice. We use $k_\text{B} = 1$.}

    \label{fig:d_versus_rho}
\end{figure}

In both the metallic and Mott insulating regimes an accurate measurement of
the double occupancy provides direct access to thermal excitations. We
analyze the crossover from thermal creation of double occupancies to
thermal depletion which is unique to a trapped system (see
Fig.~\ref{fig:d_versus_rho}). The variability of the double occupancy with
respect to temperature allows the entropy of the system to be inferred
directly by comparison with two \emph{ab-initio} theoretical methods. By
determining all other quantities entering the analysis separately and with
methods that are independent of the double occupancy measurement, we
demonstrate the versatility of the double occupancy in quantifying the
state of the system.

To obtain a quantum degenerate Fermi gas we adhere to the procedure
described in previous work \cite{Jordens2008}. A balanced spin mixture of
$^{40}\text{K}$ atoms in the $m_F=-9/2$ and $-5/2$ magnetic sublevels of
the $F=9/2$ hyperfine manifold is evaporatively cooled in a crossed beam
optical dipole trap, with less than $1.2\%$ of the atoms remaining in the
$m_F=-7/2$ state. We prepare Fermi gases with total atom numbers between
$N=30\times10^3$ and $300\times10^3$. The atom number is calibrated using
strong saturation imaging \cite{Reinaudi2007} at high magnetic field, with
a systematic error $\sim10\%$.

The optical lattice potential is then ramped up in $0.2\,\text{s}$ and has
a simple cubic symmetry with lattice constant $d=532\,\text{nm}$. Its depth
is determined from Raman-Nath diffraction of $^{87}$Rb and confirmed by
resonant excitation of atoms to higher bands \cite{Morsch2006}.  In the
lowest Bloch band the tunneling matrix element is $t/h=174(30)\,\text{Hz}$
where $h$ is Planck's constant. From transverse oscillations in the
standing wave potentials of each lattice beam and the dipole trap we
extract overall harmonic trapping frequencies of $\omega_{x,y,z}/2\pi
= [49.4(9),52.6(6),133.0(10)]\,\text{Hz}$ and a geometric mean of
$\omega/2\pi=70.1(5)\,\text{Hz}$.  The characteristic filling is
$\rho=N/N_0$ \cite{Rigol2003} where $N_0 = (12t/m\omega^2d^2)^{3/2}$ and
$m$ is the $^{40}$K atom mass.  The characteristic atom number $N_0$
relates the bandwidth to the inhomogeneity and corresponds to the atom
number per spin that yields half filling in the center of the trap at zero
temperature without interaction.

We tune the interaction between the atoms by adjusting the scattering
length in the proximity of an $s$-wave Feshbach resonance before loading
into the lattice. To determine the scattering length, the width of the
resonance was measured by using the suppressed dephasing of Bloch
oscillations \cite{Gustavsson2008} to locate the zero crossing of the
scattering length.  We obtain a width of the resonance of
$7.5(1)\,\text{G}$ which deviates from the result of Ref.~\cite{Regal2003a}
where the mean-field energy was measured. We infer the on-site interaction
energy $U$ from the scattering length and the Wannier function in the
lowest Bloch band \cite{Jaksch1998}.  This \emph{ab initio} $U$ is
experimentally validated using resonant excitation of double occupancies by
lattice modulation \cite{Jordens2008, Strohmaier2010}. We cover the range
from weak repulsion in the metallic regime to strong repulsion with a Mott
insulating core using scattering lengths between $200a_0$ and $650a_0$,
where $a_0$ is the Bohr radius.  We choose values of the Hubbard parameter
$U/6t = 1.4(2)$, $2.4(4)$, $3.2(5)$ and $4.1(7)$.  Because of the lattice
loading process, beam intensity noise and incoherent photon scattering, the
atoms heat up during preparation.  Before loading into the lattice, the
temperature in the dipole trap is around $0.13T_\text{F}$ independent of
the atom number as determined from the momentum distribution after
time-of-flight. Here $T_\text{F}$ is the Fermi temperature.  This
corresponds to an entropy per atom of $s=S/N\approx1.3$ \cite{Carr2004}.
Since the system is isolated from the environment, the temperature changes
significantly even when adiabatically loading into the lattice.  The
entropy, however only changes due to nonadiabatic processes.  Therefore we
can find a typical upper limit of the specific entropy in the lattice by
reversing the loading sequence and subsequently measuring the temperature.
Here we obtain $s<2.5$.

After loading the atoms into the lattice we determine the double occupancy.
A sudden increase of the lattice depth suppresses further tunneling. The
fraction of atoms on doubly occupied lattice sites $D$ is then obtained by
combining rf spectroscopy, Stern-Gerlach separation of the spin components
and absorption imaging \cite{Jordens2008, Strohmaier2010}. Here we account
for the independently determined offset due to the imperfection of the
initial spin mixture.  From long term reproducibility and comparison with
the adiabatic formation of molecules via magnetic field sweeps we conclude
that the relative systematic uncertainty of the double occupancy
measurement is $10\%$.

Because of the harmonic trapping potential, the temperature behavior of the
double occupancy can be markedly different from that of homogeneous systems
\cite{Koehl2006, DeLeo2008}.  In a homogeneous system the double
occupancy increases with temperature in most regimes of filling and
interaction strength due to thermal activation.  However, in a harmonically
trapped system an increase in temperature allows the atoms to reach outer
regions of the trap, in turn reducing the density in the central region: in
this case thermal excitations do not populate doubly occupied states but
rather deplete them through the decrease of the density.  The regimes
depicted in Fig.~\ref{fig:d_versus_rho} demonstrate the competition between
thermal activation and the effect of the trapping potential on the double
occupancy as a function of filling and entropy.  To extract the entropy the
experimental data are compared with theoretical results.  The curves in
Fig.~\ref{fig:d_versus_rho} correspond to the best fitting entropy and
its experimental bounds.

We apply a high-temperature series expansion \cite{Oitmaa2006} as well as
single-site dynamical mean-field theory (DMFT) \cite{Georges1996} with
a continuous time quantum Monte Carlo solver \cite{Werner2006}.  In the
experimentally relevant regime we find the high-temperature series and DMFT
to be in agreement to within $0.2\%$.  For simplicity, the theoretical
curves shown in this Letter are therefore generated using the second order
high-temperature series unless stated otherwise. The entropy is determined
from a one-parameter least-squares fit of the high-temperature series $D(s,
\rho_i)$ to the experimental data points $D_i$ weighting them according to
their statistical errors $\sigma_{Di}$. The fit minimizes
$\chi^2=\sum_i(D(s, \rho_i)-D_i)^2/\sigma_{Di}^2$.  The series is able to
accurately reproduce the measured double occupancy for all shown
interaction strengths.  We find deviations of the experimental data only at
the lowest fillings at low repulsion, indicating that for very small atom
numbers and weak interaction the cooling and loading procedure may fail to
produce a constant entropy per atom.

The size and direction of the corridors between the initial and the final
entropy in the dipole trap has implications for the usefulness of the
double occupancy when performing thermometry.  In
Fig.~\ref{fig:d_versus_rho}(a) the behavior for low repulsion $U/6t=1.4$
is shown. With increasing filling the system transforms from a dilute gas
to an increasingly dense metal with high double occupancy. In this case the
effect of the trapping potential dominates and $D$ decreases with
increasing entropy.  Because of its large $|\partial D/\partial s|$, the
regime of Fig.~\ref{fig:d_versus_rho}(a) is well suited for thermometry.

At intermediate repulsion strengths in Fig.~\ref{fig:d_versus_rho}(b) and
(c), double occupancies become increasingly suppressed and $|\partial
D/\partial s|$ decreases. For each interaction strength $\partial
D/\partial s$ changes sign at a certain filling.  These points mark the
crossover to thermal suppression of double occupancies.  If $\partial
D/\partial s$ vanishes, the theory becomes parameter-free to first order
and can be used to further determine other calibration factors, e.g., the
characteristic filling.

Fig.~\ref{fig:d_versus_rho}(d) shows data for clouds in the Mott
insulating regime. It exhibits a pronounced plateau of suppressed double
occupancy at intermediate fillings owing to a vanishing core
compressibility, a characteristic signature of a Mott insulating core
\cite{Scarola2009,Jordens2008}.  Large filling can increase the chemical
potential to values comparable with $U$ and thus create double occupancy.
In this regime the thermal activation of double occupancies dominates
over the thermal decrease of density due to the trapping potential.  If
$\partial D/\partial s > 0$ a large fraction of particles resides in the
Mott insulating core.  Here the chemical potential is high enough to
prevent holes from entering the center and additionally the density of
states is sufficiently gapped to allow only few thermally excited double
occupancies.

\begin{table}[!htb]
  \setlength{\tabcolsep}{6pt}
  \begin{tabular}{rrrrr}
     $U/6t$     & $1.4$     & $2.4$     & $3.2$     & $4.1$   \\
     \hline\hline
     $\delta_t\partial s_\text{fit}/\partial t$
                & $-0.01$   & $0.01$    & $-0.11$   & $-0.08$ \\
     $\delta_U\partial s_\text{fit}/\partial U$
                & $\sim 0$  & $-0.04$   & $0.07$    & $0.09$  \\
     $\delta_\omega\partial s_\text{fit}/\partial \omega$
                & $0.01$    & $0.07$    & $-0.07$   & $-0.07$ \\
     $\delta_N\partial s_\text{fit}/\partial N$
                & $0.06$    & $0.30$    & $-0.32$   & $-0.32$ \\
     $\delta_D\partial s_\text{fit}/\partial D$
                & $-0.16$   & $-0.30$   & $0.13$    & $0.13$  \\
     $\sigma_s$
                & $0.01$    & $0.12$    & $0.18$    & $0.07$  \\
     \hline
     total $s$  & $2.2(2)$  & $2.0(5)$  & $1.9(4)$  & $1.6(4)$
  \end{tabular}

  \caption{Error budget of the entropy determination. The table lists the
  sensitivity of the fit $\partial s_\text{fit}/\partial (\cdot)$ to the
  changes in the system's parameters scaled by their systematic errors
  $\delta_{(\cdot)}$.  For a positive contribution an increase in the
  parameter would lead to an increase in the apparent entropy. The
  contributions are added in quadrature to the fit error estimate
  $\sigma_s^2=2(\partial^2 \chi^2/\partial s^2)^{-1}$ to obtain the total
  uncertainty of the entropy.}

  \label{tab:syserrors}
\end{table}

We now consider the systematic errors of all parameters and measurements to
assess the absolute reliability of the present method in determining the
entropy.  Table~\ref{tab:syserrors} lists the contributions. The
sensitivity of the least-squares fit to variation of the respective
parameter shows the sign of the influence as well as the magnitude.  The
total relative uncertainties are below $25\%$ for all four interaction
strengths which confirms the validity of the determined entropies.  It is
apparent that the systematic errors dominate and that especially the atom
number and double occupancy calibrations are critical.  The observed
increase of the specific entropy with decreasing interaction can be
explained by an interaction-dependent global adiabaticity of the
preparation \cite{Hung2010} or by a combination of systematic errors in $N$
and $D$.

\begin{figure}[!htb]
    \includegraphics{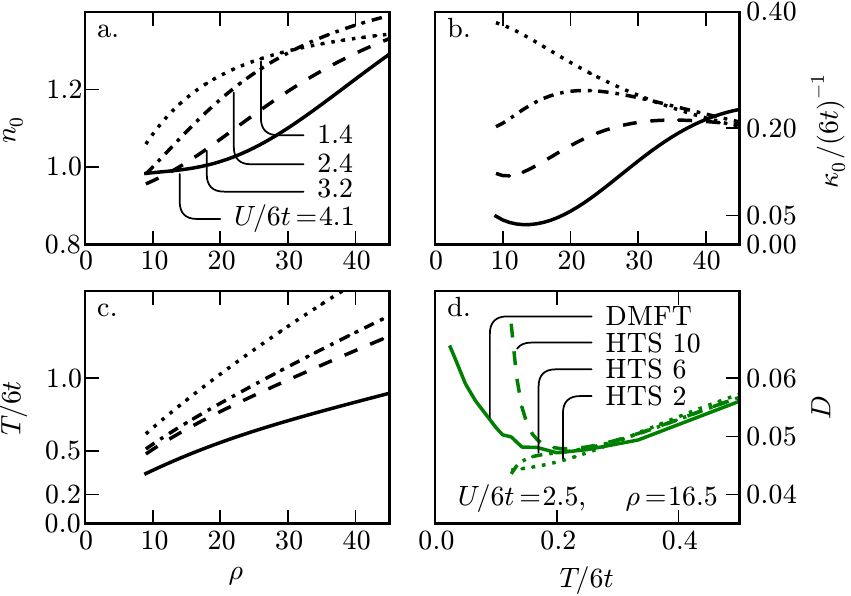}

    \caption{Properties of experimental regimes and validity of theoretical
    methods.  Panels (a)--(c) show the central density $n_0$, central
    compressibility $\kappa_0$ and temperature $T$ reached in the
    corresponding Hubbard model as a function of characteristic filling
    $\rho$ for the parameters of Fig.~\ref{fig:d_versus_rho}.  Panel (d):
    Agreement between high-temperature series (HTS, second order dotted
    line, sixth order dash-dotted, tenth order dashed line) and DMFT
    (solid line) for $U/6t=2.5$ and $\rho=16.5$ as a function of
    temperature in the lattice $T/6t$. For low temperatures $T \lesssim t$
    the series starts to diverge.}

    \label{fig:dens_compr}
\end{figure}

From the theoretical description, several unique properties of trapped
repulsively interacting Fermi-Hubbard systems can be derived.
Figures~\ref{fig:dens_compr}(a) and (b) show the central density $n_0$ and
compressibility $\kappa_0=\partial n_0/\partial\mu$ versus characteristic
filling for the interaction strengths and specific entropies of
Fig.~\ref{fig:d_versus_rho}.  The plateau in $n_0$ and the reduction of
$\kappa_0$ for $U/6t=4.1$ are signatures of the Mott insulating regime
\cite{DeLeo2008}. Compared to the result for a noninteracting system where
the ground state has a compressibility of $1.69/6t$ at half filling, the
compressibility is suppressed by a factor of 50 to values as low as
$\kappa_0\approx0.03/6t$.

The entropy as determined above needs to be related to a temperature to
allow for comparison with models of homogeneous systems.
Figure~\ref{fig:dens_compr}(c) shows this temperature in units of the half
bandwidth as a function of characteristic filling.  The behavior is
similar to that of a Fermi gas in a harmonic trap where the temperature at
constant specific entropy increases with the atom number \cite{Carr2004,
Koehl2006}.  At the lowest fillings of $\rho=5$ the temperature in the
lattice even approaches the energy scale of the tunneling $T\sim t$. At
these low temperatures the results of high-temperature series and DMFT
start to deviate considerably, see Fig.~\ref{fig:dens_compr}(d).

\begin{figure}[!htb]
    \includegraphics{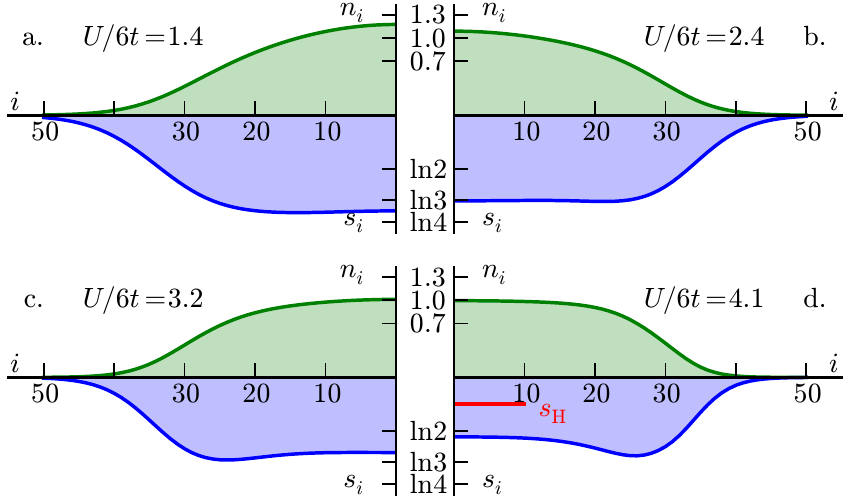}

    \caption{Density and entropy distribution in the trap. For the
    interaction strengths and entropies of Fig.~\ref{fig:d_versus_rho} the
    density $n_i$ (upward) and entropy $s_i$ (downward) per site $i$ at
    a characteristic filling of $\rho=15$ in a spherically symmetric
    system are shown. The buffering effect of the low-density shell around
    the Mott insulating core becomes clearly visible for $U/6t=4.1$. There,
    the entropy reaches values of about twice the critical entropy of the
    Heisenberg model $s_\text{H}\approx\ln2/2$.}

    \label{fig:s_versus_x}
\end{figure}

In a finite trapped system, antiferromagnetic order requires two conditions
to be met. It can only be established in regions of sufficiently constant
atom density and low specific entropy.  The experimental situation with
respect to those conditions is depicted in Fig.~\ref{fig:s_versus_x} for
a characteristic filling of $\rho=15$ and the same parameters as those
shown in Fig.~\ref{fig:d_versus_rho}.  The upward axes show the spatial
density distribution.  At low repulsive interaction
[Figs.~\ref{fig:s_versus_x}(a) and \ref{fig:s_versus_x}(b)] the system has
a density above one in the center which corresponds to a significant
doping. In Fig.~\ref{fig:s_versus_x}(c), double occupancies and holes
compensate in the center and lead to an average density close to unity
which then starts to deviate a few sites away from the center. Only for the
largest repulsion Fig.~\ref{fig:s_versus_x}(d), the Mott insulating core is
robust against the confining potential over the central 20 sites.  Here,
the density changes by less than $1\%$.

The entropy per site is shown on the downward axes in
Fig.~\ref{fig:s_versus_x}. It is highest in regions of the trap where the
density is most variable. For small interaction strengths it reaches values
close to the maximum of $s_i=\ln4$ in some regions.  For large repulsion
the sites in the perimeter of the cloud where $n_i\sim2/3$ carry most of
the entropy.  This can be understood in the atomic limit at large $U$.  At
$n_i=2/3$ each site has three equally likely states and can accommodate
$s_i=\ln3$ of entropy. The Mott insulating core can only absorb $\ln2$ of
spin entropy. Mean-field theory of the Heisenberg model predicts this to
coincide with the entropy at its critical point.  However, quantum
fluctuations lower the entropy at the N\'eel temperature where magnetic
long-range order sets in to about $s_\text{H} \approx \ln2/2 \approx 0.35$
\cite{Werner2005}. We have verified numerically that for the Heisenberg
model with exchange coupling $J$ the entropy is $s = 0.338(5)$ at
$T_\text{N\'eel}/J = 0.946(1)$ \cite{Sandvik1998,Wessel2010} by integrating
the energy, $S = \int dE/T$. Integration from above or below the N\'eel
temperature agree.  Additionally, we performed a new study of the Hubbard
model using a diagrammatic determinant Monte Carlo method
\cite{Burovski2008}. For $U/t = 8$, the critical temperature is
$T_\text{N\'eel}/t = 0.325(7)$ \footnote{This estimate improves on the
earlier result $T_\text{N\'eel}/t=0.31(1)$ of \cite{Staudt2000}}, and the
critical entropy $s_\text{N\'eel} = 0.345(45)$. This differs from the
mean-field calculation including fluctuation corrections of
Ref.~\cite{Koetsier2008} and is a factor of 2 less than the experimental
results presented here.

We acknowledge funding from SNF, NCCR MaNEP, NAME-QUAM (EU, FET-Open),
SCALA (EU), ANR (FABIOLA and FAMOUS), DARPA-OLE, and ``Triangle de la
Physique''.

\bibliographystyle{prl}
\bibliography{fmi}{}

\begin{thebibliography}{10}
\providecommand{\bibnamefont}[1]{#1}
\providecommand{\bibfnamefont}[1]{#1}
\providecommand{\bibinfo}[2]{#2}
\providecommand{\biburlhref}[2]{\href{#1}{#2}}

\bibitem{DeMarco1999}
\biburlhref{http://dx.doi.org/10.1126/science.285.5434.1703}{%
\bibinfo{author}{\bibfnamefont{B.}~\bibnamefont{{D}e{M}arco}} \bibnamefont{and}
  \bibinfo{author}{\bibfnamefont{D.~S.} \bibnamefont{{J}in}},
  \bibinfo{journal}{{S}cience} \textbf{\bibinfo{volume}{285}},
  \bibinfo{pages}{1703} (\bibinfo{year}{1999}).}

\bibitem{Loftus2002}
\biburlhref{http://dx.doi.org/10.1103/PhysRevLett.88.173201}{%
\bibinfo{author}{\bibfnamefont{T.}~\bibnamefont{{L}oftus}} \emph{et~al.},
  \bibinfo{journal}{{P}hys. {R}ev. {L}ett.} \textbf{\bibinfo{volume}{88}},
  \bibinfo{pages}{173201} (\bibinfo{year}{2002}).}

\bibitem{Greiner2002}
\biburlhref{http://dx.doi.org/10.1038/415039a}{%
\bibinfo{author}{\bibfnamefont{M.}~\bibnamefont{{G}reiner}} \emph{et~al.},
  \bibinfo{journal}{{N}ature ({L}ondon)} \textbf{\bibinfo{volume}{415}},
  \bibinfo{pages}{39} (\bibinfo{year}{2002}).}

\bibitem{Koehl2005}
\biburlhref{http://dx.doi.org/10.1103/PhysRevLett.94.080403}{%
\bibinfo{author}{\bibfnamefont{M.}~\bibnamefont{{K}öhl}} \emph{et~al.},
  \bibinfo{journal}{{P}hys. {R}ev. {L}ett.} \textbf{\bibinfo{volume}{94}},
  \bibinfo{pages}{080403} (\bibinfo{year}{2005}).}

\bibitem{Hofstetter2002}
\biburlhref{http://dx.doi.org/10.1103/PhysRevLett.89.220407}{%
\bibinfo{author}{\bibfnamefont{W.}~\bibnamefont{{H}ofstetter}} \emph{et~al.},
  \bibinfo{journal}{{P}hys. {R}ev. {L}ett.} \textbf{\bibinfo{volume}{89}},
  \bibinfo{pages}{220407} (\bibinfo{year}{2002}).}

\bibitem{Jordens2008}
\biburlhref{http://dx.doi.org/10.1038/nature07244}{%
\bibinfo{author}{\bibfnamefont{R.}~\bibnamefont{{J}ördens}} \emph{et~al.},
  \bibinfo{journal}{{N}ature ({L}ondon)} \textbf{\bibinfo{volume}{455}},
  \bibinfo{pages}{204} (\bibinfo{year}{2008}).}

\bibitem{Schneider2008}
\biburlhref{http://dx.doi.org/10.1126/science.1165449}{%
\bibinfo{author}{\bibfnamefont{U.}~\bibnamefont{{S}chneider}} \emph{et~al.},
  \bibinfo{journal}{{S}cience} \textbf{\bibinfo{volume}{322}},
  \bibinfo{pages}{1520} (\bibinfo{year}{2008}).}

\bibitem{Trotzky2009}
\biburlhref{http://arxiv.org/abs/0905.4882}{%
\bibinfo{author}{\bibfnamefont{S.}~\bibnamefont{{T}rotzky}} \emph{et~al.},
  \bibinfo{journal}{ar{X}iv:0905.4882v1}  (\bibinfo{year}{2009}).}

\bibitem{Weld2009}
\biburlhref{http://dx.doi.org/10.1103/PhysRevLett.103.245301}{%
\bibinfo{author}{\bibfnamefont{D.~M.} \bibnamefont{{W}eld}} \emph{et~al.},
  \bibinfo{journal}{{P}hys. {R}ev. {L}ett.} \textbf{\bibinfo{volume}{103}},
  \bibinfo{eid}{245301} (\bibinfo{year}{2009}).}

\bibitem{Koehl2006}
\biburlhref{http://dx.doi.org/10.1103/PhysRevA.73.031601}{%
\bibinfo{author}{\bibfnamefont{M.}~\bibnamefont{{K}öhl}},
  \bibinfo{journal}{{P}hys. {R}ev. {A}} \textbf{\bibinfo{volume}{73}},
  \bibinfo{pages}{031601} (\bibinfo{year}{2006}).}

\bibitem{Strohmaier2007}
\biburlhref{http://dx.doi.org/10.1103/PhysRevLett.99.220601}{%
\bibinfo{author}{\bibfnamefont{N.}~\bibnamefont{{S}trohmaier}} \emph{et~al.},
  \bibinfo{journal}{{P}hys. {R}ev. {L}ett.} \textbf{\bibinfo{volume}{99}},
  \bibinfo{eid}{220601} (\bibinfo{year}{2007}).}

\bibitem{Scarola2009}
\biburlhref{http://dx.doi.org/10.1103/PhysRevLett.102.135302}{%
\bibinfo{author}{\bibfnamefont{V.~W.} \bibnamefont{{S}carola}} \emph{et~al.},
  \bibinfo{journal}{{P}hys. {R}ev. {L}ett.} \textbf{\bibinfo{volume}{102}},
  \bibinfo{eid}{135302} (\bibinfo{year}{2009}).}

\bibitem{Reinaudi2007}
\biburlhref{http://dx.doi.org/10.1364/OL.32.003143}{%
\bibinfo{author}{\bibfnamefont{G.}~\bibnamefont{{R}einaudi}} \emph{et~al.},
  \bibinfo{journal}{{O}pt. {L}ett.} \textbf{\bibinfo{volume}{32}},
  \bibinfo{pages}{3143} (\bibinfo{year}{2007}).}

\bibitem{Morsch2006}
\biburlhref{http://dx.doi.org/10.1103/RevModPhys.78.179}{%
\bibinfo{author}{\bibfnamefont{O.}~\bibnamefont{{M}orsch}} \bibnamefont{and}
  \bibinfo{author}{\bibfnamefont{M.}~\bibnamefont{{O}berthaler}},
  \bibinfo{journal}{{R}ev. {M}od. {P}hys.} \textbf{\bibinfo{volume}{78}},
  \bibinfo{pages}{179} (\bibinfo{year}{2006}).}

\bibitem{Rigol2003}
\biburlhref{http://dx.doi.org/10.1103/PhysRevLett.91.130403}{%
\bibinfo{author}{\bibfnamefont{M.}~\bibnamefont{{R}igol}} \emph{et~al.},
  \bibinfo{journal}{{P}hys. {R}ev. {L}ett.} \textbf{\bibinfo{volume}{91}},
  \bibinfo{pages}{130403} (\bibinfo{year}{2003}).}

\bibitem{Gustavsson2008}
\biburlhref{http://dx.doi.org/10.1103/PhysRevLett.100.080404}{%
\bibinfo{author}{\bibfnamefont{M.}~\bibnamefont{{G}ustavsson}} \emph{et~al.},
  \bibinfo{journal}{{P}hys. {R}ev. {L}ett.} \textbf{\bibinfo{volume}{100}},
  \bibinfo{eid}{080404} (\bibinfo{year}{2008}).}

\bibitem{Regal2003a}
\biburlhref{http://dx.doi.org/10.1103/PhysRevLett.90.230404}{%
\bibinfo{author}{\bibfnamefont{C.~A.} \bibnamefont{{R}egal}} \bibnamefont{and}
  \bibinfo{author}{\bibfnamefont{D.~S.} \bibnamefont{{J}in}},
  \bibinfo{journal}{{P}hys. {R}ev. {L}ett.} \textbf{\bibinfo{volume}{90}},
  \bibinfo{pages}{230404} (\bibinfo{year}{2003}).}

\bibitem{Jaksch1998}
\biburlhref{http://dx.doi.org/10.1103/PhysRevLett.81.3108}{%
\bibinfo{author}{\bibfnamefont{D.}~\bibnamefont{{J}aksch}} \emph{et~al.},
  \bibinfo{journal}{{P}hys. {R}ev. {L}ett.} \textbf{\bibinfo{volume}{81}},
  \bibinfo{pages}{3108} (\bibinfo{year}{1998}).}

\bibitem{Strohmaier2010}
\biburlhref{http://dx.doi.org/10.1103/PhysRevLett.104.080401}{%
\bibinfo{author}{\bibfnamefont{N.}~\bibnamefont{{S}trohmaier}} \emph{et~al.},
  \bibinfo{journal}{{P}hys. {R}ev. {L}ett.} \textbf{\bibinfo{volume}{104}},
  \bibinfo{pages}{080401} (\bibinfo{year}{2010}).}

\bibitem{Carr2004}
\biburlhref{http://dx.doi.org/10.1103/PhysRevLett.92.150404}{%
\bibinfo{author}{\bibfnamefont{L.~D.} \bibnamefont{{C}arr}} \emph{et~al.},
  \bibinfo{journal}{{P}hys. {R}ev. {L}ett.} \textbf{\bibinfo{volume}{92}},
  \bibinfo{pages}{150404} (\bibinfo{year}{2004}).}

\bibitem{DeLeo2008}
\biburlhref{http://dx.doi.org/10.1103/PhysRevLett.101.210403}{%
\bibinfo{author}{\bibfnamefont{L.}~\bibnamefont{{D}e {L}eo}} \emph{et~al.},
  \bibinfo{journal}{{P}hys. {R}ev. {L}ett.} \textbf{\bibinfo{volume}{101}},
  \bibinfo{eid}{210403} (\bibinfo{year}{2008}).}

\bibitem{Oitmaa2006}

\bibinfo{author}{\bibfnamefont{J.}~\bibnamefont{{O}itmaa}} \emph{et~al.},
  \emph{\bibinfo{title}{{S}eries expansion methods for strongly interacting
  lattice models}} (\bibinfo{publisher}{Cambridge University Press},
  \bibinfo{address}{Cambridge}, \bibinfo{year}{2006}).

\bibitem{Georges1996}
\biburlhref{http://dx.doi.org/10.1103/RevModPhys.68.13}{%
\bibinfo{author}{\bibfnamefont{A.}~\bibnamefont{{G}eorges}} \emph{et~al.},
  \bibinfo{journal}{{R}ev. {M}od. {P}hys.} \textbf{\bibinfo{volume}{68}},
  \bibinfo{pages}{13} (\bibinfo{year}{1996}).}

\bibitem{Werner2006}
\biburlhref{http://dx.doi.org/10.1103/PhysRevLett.97.076405}{%
\bibinfo{author}{\bibfnamefont{P.}~\bibnamefont{{W}erner}} \emph{et~al.},
  \bibinfo{journal}{{P}hys. {R}ev. {L}ett.} \textbf{\bibinfo{volume}{97}},
  \bibinfo{eid}{076405} (\bibinfo{year}{2006}).}

\bibitem{Hung2010}
\biburlhref{http://arxiv.org/abs/1003.0855}{%
\bibinfo{author}{\bibfnamefont{C.-L.} \bibnamefont{{H}ung}} \emph{et~al.},
  \bibinfo{journal}{ar{X}iv:0910.1382v1}  (\bibinfo{year}{2010}).}

\bibitem{Werner2005}
\biburlhref{http://dx.doi.org/10.1103/PhysRevLett.95.056401}{%
\bibinfo{author}{\bibfnamefont{F.}~\bibnamefont{{W}erner}} \emph{et~al.},
  \bibinfo{journal}{{P}hys. {R}ev. {L}ett.} \textbf{\bibinfo{volume}{95}},
  \bibinfo{pages}{056401} (\bibinfo{year}{2005}).}

\bibitem{Sandvik1998}
\biburlhref{http://dx.doi.org/10.1103/PhysRevLett.80.5196}{%
\bibinfo{author}{\bibfnamefont{A.~W.} \bibnamefont{{S}andvik}},
  \bibinfo{journal}{{P}hys. {R}ev. {L}ett.} \textbf{\bibinfo{volume}{80}},
  \bibinfo{pages}{5196} (\bibinfo{year}{1998}).}

\bibitem{Wessel2010}
\biburlhref{http://dx.doi.org/10.1103/PhysRevB.81.052405}{%
\bibinfo{author}{\bibfnamefont{S.}~\bibnamefont{{W}essel}},
  \bibinfo{journal}{{P}hys. {R}ev. {B}} \textbf{\bibinfo{volume}{81}},
  \bibinfo{pages}{052405} (\bibinfo{year}{2010}).}

\bibitem{Burovski2008}
\biburlhref{http://dx.doi.org/10.1103/PhysRevLett.101.090402}{%
\bibinfo{author}{\bibfnamefont{E.}~\bibnamefont{{B}urovski}} \emph{et~al.},
  \bibinfo{journal}{{P}hys. {R}ev. {L}ett.} \textbf{\bibinfo{volume}{101}},
  \bibinfo{eid}{090402} (\bibinfo{year}{2008}).}

\bibitem{Koetsier2008}
\biburlhref{http://dx.doi.org/10.1103/PhysRevA.77.023623}{%
\bibinfo{author}{\bibfnamefont{A.}~\bibnamefont{{K}oetsier}} \emph{et~al.},
  \bibinfo{journal}{{P}hys. {R}ev. {A}} \textbf{\bibinfo{volume}{77}},
  \bibinfo{eid}{023623} (\bibinfo{year}{2008}).}

\bibitem{Staudt2000}
\biburlhref{http://dx.doi.org/10.1007/s100510070120}{%
\bibinfo{author}{\bibfnamefont{R.}~\bibnamefont{{S}taudt}} \emph{et~al.},
  \bibinfo{journal}{{E}ur. {P}hys. {J}. {B}} \textbf{\bibinfo{volume}{17}},
  \bibinfo{pages}{411} (\bibinfo{year}{2000}).}

\end{thebibliography}

\end{document}